\documentclass[journal,twoside,web]{ieeecolor}
\usepackage{cite}
\providecommand{\refname}{References}
\usepackage{generic}
\usepackage{amsmath,amssymb,amsfonts}
\usepackage{graphicx}
\usepackage{algorithm,algorithmic}
\usepackage{xspace}
\usepackage{textcomp}
\usepackage{array}

\def\BibTeX{{\rm B\kern-.05em{\sc i\kern-.025em b}\kern-.08em
    T\kern-.1667em\lower.7ex\hbox{E}\kern-.125emX}}
\begin{document}

\title{EffortNet: A Deep Learning Framework for Objective Assessment of Speech Enhancement Technologies Using EEG-Based Alpha Oscillations}

\author{Ching-Chih Sung$^{*}$~\IEEEmembership{Student Member,~IEEE}, Cheng-Hung Hsin$^{*}$, Yu-Anne Shiah, Bo-Jyun Lin, Yi-Xuan Lai, Chia-Ying Lee, Yu-Te Wang, Borchin Su, and Yu Tsao~\IEEEmembership{Senior Member,~IEEE}
\thanks{Ching-Chih Sung is with the Graduate Institute of Communication Engineering, National Taiwan University, Taipei, Taiwan (e-mail: d07942007@ntu.edu.tw)}
\thanks{Cheng-Hung Hsin was with the Research Center for Information Technology Innovation, Academia Sinica, Taipei, Taiwan, and is with the Department of Foreign Languages and Literatures, National Chung Hsing University, Taichung, Taiwan (email: chhsin@nchu.edu.tw)}
\thanks{Yu-Anne Shiah is with the Research Center for Information Technology Innovation, Academia Sinica, Taipei, Taiwan (email: anneshiah@citi.sinica.edu.tw)}
\thanks{Bo-Jyun Lin is with the Department of Mechanical Engineering, National Taiwan University, Taipei, Taiwan, and also with the Research Center for Information Technology Innovation, Academia Sinica, Taipei, Taiwan (e-mail: b11502064@ntu.edu.tw)}
\thanks{Yi-Xuan Lai is with the Department of Computer Science, National Yang Ming Chiao Tung University, Hsinchu, Taiwan; (e-mail: doralai0721.mg10@nycu.edu.tw)}
\thanks{Chia-Ying Lee is with the Institute of Linguistics, Academia Sinica, Taipei, Taiwan; also with the Institute of Cognitive Neuroscience, National Central University, Taoyuan, Taiwan; and with the Research Center for Mind, Brain, and Learning, National Chengchi University, Taipei, Taiwan (e-mail: chiaying@gate.sinica.edu.tw)}
\thanks{Yu-Te Wang is with the Research Center for Information Technology Innovation, Academia Sinica, Taipei, Taiwan (email: yutewang@citi.sinica.edu.tw)}
\thanks{Borching Su is with the Graduate Institute of Communication Engineering, National Taiwan University, Taipei, Taiwan, (email: borching@ntu.edu.tw )}
\thanks{Yu Tsao is with the Research Center for Information Technology Innovation, Academia Sinica, Taipei, Taiwan, and is also with the Department of Electrical Engineering, Chung Yuan Christian University, Taoyuan, Taiwan (e-mail: yu.tsao@citi.sinica.edu.tw)}
\thanks{*These authors contributed equally to this work.}}

\maketitle

\begin{abstract}
This paper presents EffortNet, a novel deep learning framework for decoding individual listening effort from electroencephalography (EEG) during speech comprehension. Listening effort represents a significant challenge in speech-hearing research, particularly for aging populations and those with hearing impairment. We collected 64-channel EEG data from 122 participants during speech comprehension under four conditions: clean, noisy, MMSE-enhanced, and Transformer-enhanced speech. Statistical analyses confirmed that alpha oscillations (8-13 Hz) exhibited significantly higher power during noisy speech processing compared to clean or enhanced conditions, confirming their validity as objective biomarkers of listening effort. To address the substantial inter-individual variability in EEG signals, EffortNet integrates three complementary learning paradigms: self-supervised learning to leverage unlabeled data, incremental learning for progressive adaptation to individual characteristics, and transfer learning for efficient knowledge transfer to new subjects. Our experimental results demonstrate that EffortNet achieves $80.9$\% classification accuracy with only $40$\% training data from new subjects, significantly outperforming conventional CNN ($62.3$\%) and STAnet ($61.1$\%) models. The probability-based metric derived from our model revealed that Transformer-enhanced speech elicited neural responses more similar to clean speech than MMSE-enhanced speech. This finding contrasted with subjective intelligibility ratings but aligned with objective metrics. The proposed framework provides a practical solution for personalized assessment of hearing technologies, with implications for designing cognitive-aware speech enhancement systems.

\end{abstract}


\section{Introduction}
\label{sec:introduction}

\IEEEPARstart{I}{n} everyday communication, people frequently encounter noise that compromises speech comprehension and increases listening effort \cite{pichora2016hearing}. This challenge is particularly significant given the aging global population. The United Nations projects that the population over 65 will increase to 2.2 billion by 2070, while currently 430 million people ($5.5$\% of the global population) have hearing loss, with age-related factors being a major cause \cite{un2024wpp, who2021hearing}. The consequences of hearing loss extend beyond communication difficulties, as it is associated with a higher risk of dementia \cite{uhlmann1989relationship,lin2011hearing,liu2019association,deal2017hearing,livingston2020dementia,yu2024adult}. While hearing aids may slow cognitive decline \cite{sarant2024cochlea,sarant2024enhance,lin2023hearing} and recent regulatory changes aim to expand access (e.g., over-the-counter devices), technical limitations such as distortion and residual noise in assistive hearing devices can increase listening effort\cite{das2021fundamentals,hsin2025exploring,ohlenforst2017effects}, preventing users from developing consistent wearing habits. Moreover, current clinical assessments such as pure tone audiometry may not fully capture real-world speech understanding performance or the effort required in daily communication\cite{hind2011prevalence,ruggles2011normal}. Consequently, developing practical and reliable objective measurements for hearing assessment and listening effort is crucial for both clinical applications and evaluating assistive technologies in an aging global population.

However, accurately measuring listening effort presents significant challenges due to its complex and multidimensional nature. Current behavioral measures encompass diverse approaches, such as self-report ratings, response times, and dual-task performance, yet extensive research has revealed substantial inconsistencies among these measures themselves \cite{mcgarrigle2014listening, pichora2016hearing, shields2023exploring}. For instance, self-reported effort often fails to correlate with behavioral measures like dual-task performance or response times \cite{mcgarrigle2017measuring,mcgarrigle2017pupillometry,alhanbali2019measures}. These inconsistencies persist even when measures are collected simultaneously during the same task \cite{alhanbali2019measures, decruy2020top}. Research has found that limited numbers of correlational analyses between different listening effort measures achieved statistical significance\cite{shields2023exploring}. Furthermore, behavioral measures also show limited correlations with physiological indicators such as pupil dilation or EEG patterns \cite{kestens2023p300,alhanbali2019measures,mcgarrigle2017measuring,mcgarrigle2017pupillometry}. This suggests that different measures may tap into distinct dimensions of listening effort, such as cognitive load, arousal, or motivational engagement, rather than reflecting a unified construct \cite{mcgarrigle2014listening,francis2020listening,pichora2016hearing}.

While speech enhancement technologies aim to mitigate noise effects, evaluating their effectiveness is complicated by additional measurement challenges \cite{hsin2025exploring,ohlenforst2017effects}. The field of acoustic engineering has established automatic metrics such as Perceptual Evaluation of Speech Quality (PESQ) for quality assessment \cite{rix2001perceptual} and Short-Time Objective Intelligibility (STOI) for intelligibility prediction \cite{taal2011algorithm}. The former models human auditory perception while the latter analyzes temporal envelope correlations between clean and degraded speech across frequency bands. However, despite these objective metrics often suggesting improvements with enhanced speech, human listeners frequently rate unprocessed noisy speech higher in both quality and intelligibility, with limited correlations between objective metrics and subjective ratings \cite{zezario2022deep, rosenbaum2023differentiable, manjunath2009limitations,hines2013robustness,chen2021inqss,lo2019mosnet, cooper2024review}. This discrepancy between automatic metrics and behavioral assessments, combined with the inconsistencies within behavioral measures themselves, underscores the fundamental difficulty in capturing listening effort through traditional measurement approaches.

Given these fundamental measurement challenges, electroencephalography (EEG) has emerged as a particularly valuable approach for obtaining objective biomarkers of speech processing\cite{francis2020listening,mcgarrigle2014listening,ohlenforst2017effects, pichora2016hearing}. Unlike behavioral measures that aggregate effects across multiple processing stages and may not reflect identical underlying cognitive processes \cite{kemmer2004syntactic, lau2013dissociating}, neurophysiological indicators can provide more direct insights into the neural mechanisms of listening effort. Recent EEG studies have demonstrated the effectiveness of speech enhancement technologies beyond traditional assessments \cite{hsin2025exploring,slugocki2020effects,slugocki2024alpha,slugocki2024using,slugocki2025using,yu2018erp,yu2019analyzing}. For example, the event-related potential N400 has been utilized to show that enhanced speech can facilitate semantic processing\cite{hsin2025exploring}, providing neurophysiological evidence of speech enhancement benefits that might not be captured by traditional assessments.

Among various neural markers, alpha oscillations (8-13 Hz) have emerged as a particularly promising proxy for listening effort \cite{mcgarrigle2014listening,pichora2016hearing,francis2020listening}. Alpha oscillations serve as a central mechanism for auditory selective inhibition during challenging listening conditions \cite{strauss2014cortical,weisz2011alpha}, with increased power reflecting the suppression of task-irrelevant information and distractors \cite{wostmann2016spatiotemporal,wostmann2019alpha}. Studies have demonstrated that alpha activity could index speech intelligibility, with suppressed alpha power associated with successful comprehension and increased power reflecting greater listening effort \cite{becker2013left,dimitrijevic2017cortical,dimitrijevic2019neural,drijvers2016alpha,obleser2012adverse,obleser2012suppressed,paul2020poor,paul2021cortical,ryan2022impact,petersen2015hearing,slugocki2024alpha,slugocki2024using}. Recently, research has demonstrated that effective hearing aid algorithms significantly reduce alpha power during speech-in-noise tasks, providing neurophysiological evidence of alleviated listening effort beyond behavioral improvements \cite{slugocki2024alpha,slugocki2024using}. These converging findings establish alpha oscillations as robust objective biomarkers for quantifying listening effort during speech comprehension under various acoustic conditions.

Despite these promising findings, translating this knowledge into practical applications faces significant challenges \cite{pichora2016hearing}. Statistical analyses based on group-level data are insufficient for developing alpha oscillations as a functional biomarker in real-world settings. To be practically useful, there is a need for techniques that can detect and interpret differences in the alpha oscillations at the individual level, enabling personalized assessment and intervention. This individual-level application presents several obstacles: EEG data exhibit substantial inter-subject variability; recordings may contain artifacts that can mask subtle alpha modulations; and collecting high-quality EEG data requires specialized equipment and expertise, making data acquisition for clinical applications challenging.

Deep learning (DL) approaches offer promising solutions to these limitations. Recent advances in data-driven modeling techniques can account for individual variability by identifying person-specific patterns in neural activity. These methods can effectively handle noisy data by learning to distinguish signal from noise and can generate robust predictions with limited data samples. DL techniques have demonstrated remarkable efficacy in the classification of EEG data across various neurological applications \cite{xie2022transformer, hosseini2020review}. These methods can detect subtle neural signatures associated with different cognitive states and sensory processing conditions.

Among the various DL approaches, convolutional neural networks (CNNs) have received widespread recognition in EEG classification tasks \cite{zhang2019vulnerability,tao2022decoding,tang2023upper,lin2021cnn,meng2023user,peng2022tie,park2023spatio} due to their ability to automatically extract relevant spatial and temporal features from high-dimensional neurophysiological data.
In speech-hearing research, DL models have been applied to auditory spatial attention detection (ASAD) based on EEG recordings. Vandecappelle et al. \cite{vandecappelle2021eeg} demonstrated that CNNs can effectively decode the directional focus of auditory attention from frequency band information in EEG signals without requiring access to clean speech envelopes, achieving over $80$\% accuracy with only 1-second EEG windows. Research has further developed spatiotemporal attention networks that integrate both spatial and temporal attention mechanisms to improve performance in ASAD dataset\cite{su2022stanet}.

Despite these advances, a major challenge in DL-based EEG analysis is developing models that generalize effectively across different subjects due to substantial inter-individual variability in EEG signals \cite{altaheri2023deep,hosseini2020review}. Subject-dependent models achieve high accuracy but require extensive individual calibration, while cross-subject and subject-independent models typically experience significant performance degradation when confronted with diverse neural activity patterns \cite{sturm2016interpretable,pandey2022subject, zhang2023subject, kwon2019subject}. Recent improvements in subject-independent approaches include variational mode decomposition with deep neural networks \cite{pandey2022subject}, generative adversarial networks for data augmentation \cite{zhang2023subject}, and transfer learning methods \cite{ma2023cross, li2023mdtl, liu2022multi} to enhance cross-subject model performance.

Previous research has focused predominantly on alpha modulations during auditory selective attention tasks. Limited studies have examined alpha oscillations specifically as an index for listening effort and how speech enhancement technologies might mitigate this effort. To address this gap and the challenges of individual variability, we first conducted a large-scale EEG experiment with 122 participants, statistically examining alpha oscillations during speech comprehension under clean, noisy, and enhanced conditions to create this extensive dataset. This dataset provides a foundation for exploring both the neural mechanisms underlying listening effort and the effectiveness of DL methods in capturing individual-specific neural patterns.

Our approach integrates three complementary learning paradigms to develop a personalized classification framework for alpha band EEG activities: self-supervised learning (SSL) \cite{liu2021self,rafiei2022self,wang2023self}, incremental learning (IL) \cite{van2022three,deng2023centroid}, and fine-tuning (F-T) \cite{ribani2019survey,wang2023calibration,wan2021review}. SSL leverages unlabeled data through augmentation techniques to improve model robustness while reducing dependence on extensive labeled data. IL enables progressive assimilation of individual-specific features across multiple subjects, enhancing cross-individual adaptability. F-T based on the model obtained through IL allows efficient adaptation to new subjects, reducing the need for extensive new data collection.

By combining these paradigms with CNN-based models, our methodology aims to reduce the cost of data labeling and cleaning, handle inter-individual variability, and enhance model adaptability to data from new subjects for real-world applications.

This study makes three primary contributions: (1) establishing a large-scale EEG dataset for speech comprehension under clean, noisy, and enhanced conditions; (2) demonstrating the practical utility of alpha oscillations as a personalized assessment tool for listening effort; and (3) validating an ensemble framework that incorporates a pretrained model, a multi-subject foundation model, and a downstream task model with a limited amount of training data. By developing methods that can reliably detect and interpret alpha-based neural modulations at the individual level, this study seeks to advance the application of neural markers in personalized hearing assessment and speech enhancement technology evaluation.

\section{METHODS}
\label{sec:haaqi_net}

\subsection{EEG Data Collection}
This study conducted an EEG experiment to collect data from healthy young adults who were instructed to listen to speech under various acoustic conditions. Participants were presented with spoken sentences selected from a standardized corpus in four conditions (clean, noisy, and two enhanced versions) while recording 64-channel EEG responses. Participants performed comprehension and intelligibility assessment tasks during the experiment. All EEG data were carefully preprocessed for deep learning analysis of alpha oscillations.

\subsubsection{Participants}
One hundred and twenty-two healthy young adults (68 females; age range: 20-30 years, $mean$ $(M) = 24.17$, $standard$ $deviation$ $(SD) = 3.01$) participated in this EEG study investigating cognitive load during speech comprehension. All participants were right-handed native Mandarin Chinese speakers with a mean of 16.06 years of education ($SD = 1.58$). Participants reported no history of neurological or psychiatric disorders, had normal or corrected-to-normal vision, and demonstrated normal hearing thresholds $(\leq 25\,\text{dB})$ averaged across 500, 1000, 2000, and 4000 Hz in their better ear, as confirmed by pure-tone audiometry (PTA). Two participants with mild hearing loss (32.5 and 37.5 dB) were retained in the analysis, as their thresholds were substantially below the 65-dB sound pressure level (SPL) used in the experimental paradigm. The study protocol, including participant recruitment, experimental procedures, informed consent, and data privacy measures, was approved by and conducted in accordance with the guidelines of the local Institutional Review Board (IRB) at Academia Sinica (Approval No. AS-IRB-BM-24029, [2024-05-07]).

\subsubsection{Stimuli}
The experimental stimuli consisted of sentences from the Taiwanese Mandarin Hearing in Noise Test (TMHINT) \cite{wong2007development, wong2008taiwanese}. The TMHINT corpus contains phonemically and tonally balanced sentences presented in a natural conversational style without colloquial expressions, with each sentence comprising exactly 10 Chinese characters. These characteristics provided a well-controlled stimulus set for investigating neural responses to speech in noise. One male and one female native Mandarin speaker recorded 52 sentences each (104 total) in a quiet room using a 16-bit format at a 16-kHz sampling rate \cite{chen2021inqss}. The speakers maintained natural speed, intonation, and prosody during recording, with each utterance lasting approximately 3 seconds. All recordings were normalized to 65 dB SPL to approximate typical conversational volume. Speech-shaped noise (SSN) was added to all clean recordings at a signal-to-noise ratio (SNR) of +2 dB, selected on the basis of environmental studies \cite{smeds2015estimation,wu2018characteristics}. The noisy recordings were subsequently processed using Transformer-based and minimum mean square error (MMSE)-based speech enhancement models\cite{ephraim1984speech,lu2023improving}. The experimental protocol included 104 sentences equally distributed across four conditions (26 each: clean, noisy, Transformer-enhanced, and MMSE-enhanced) and balanced between speakers (13 sentences per gender in each condition).
\subsubsection{Procedure}
\begin{figure}[!t]
    \centering
    \includegraphics[width=\linewidth]{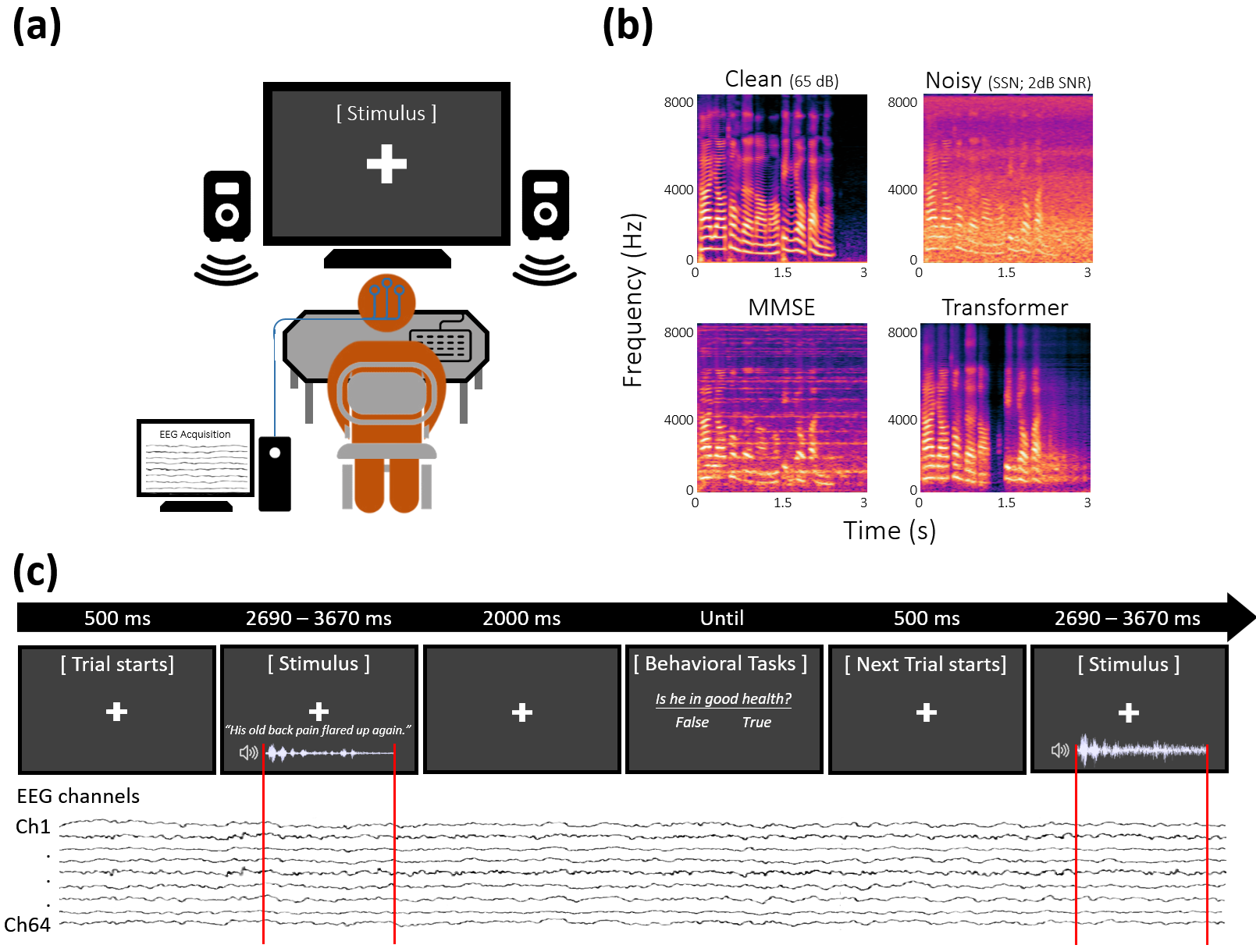}
    \caption{Experimental setup and procedure.
        \textbf{(a)} Simulated environment for EEG recording. \textbf{(b)} Spectrograms of four listening conditions: clean, noisy, MMSE-enhanced, and Transformer-enhanced speech. \textbf{(c)} Trial structure showing fixation cross during stimulus presentation and EEG recording timeline.
    }
    \label{fig:fig5}
\end{figure}

The EEG experiment was conducted in a sound-attenuated, electrically shielded room using MATLAB (R2014b) and Psychtoolbox 3. Each participant was seated 90 cm from an LCD screen, with two loudspeakers positioned at the same distance for binaural auditory stimulus presentation (see Fig.\ref{fig:fig5}). Prior to EEG recording, participants received instructions to minimize muscle and ocular artifacts during stimulus presentation and completed eight practice trials to familiarize themselves with the experimental procedures. A SPL meter verified that stimulus intensity remained at 65 dB SPL, and researchers confirmed that all participants could hear and read the stimuli clearly.

The formal experimental session consisted of 104 randomized trials divided into four blocks of approximately 6 minutes each. Stimuli were selected from the TMHINT corpus, consisting of four listening conditions: clean, noisy, MMSE-enhanced, and Transformer-enhanced speech.  Each trial began with a 500-ms fixation cross, followed by a spoken sentence presentation during fixation (2,690 to 3,670 ms). The fixation remained onscreen for an additional 2,000 ms after stimulus offset. Participants were instructed to maintain fixation on the central white cross throughout stimulus presentation. For half of the trials, comprehension was assessed through either true/false questions or forced word-choice questions (25\% each). For instance, following the sentence "His old back pain flared up again," participants might encounter either "Is he in good health? –True / False" or "Which word did you hear? - back pain / rheumatism." These two question types were counterbalanced to ensure attention to both individual words and overall sentence meaning. For the remaining trials, participants provided subjective intelligibility ratings on a scale of 1 to 4 (4 being most intelligible). Participants entered responses via keyboard after question onset to minimize EEG artifacts, and trials advanced upon response completion. EEG data were recorded continuously and subsequently epoched based on stimulus onset (see Fig. \ref{fig:fig5}).

\subsubsection{EEG Data Acquisition and Preprocessing}
EEG data were recorded using a 64-channel QuickCap system (Neuromedical Supplies) with sintered Ag/AgCl electrodes. The recording reference electrode was positioned on the scalp vertex between Cz and CPz, with the ground electrode placed anterior to Fz. Data were digitized continuously at 1,000 Hz using SynAmps2 (Neuroscan Inc.) and filtered with a 400-Hz low-pass filter for offline analysis. The recordings were subsequently re-referenced to the averaged mastoid electrodes (M1/M2). Ocular activity was monitored using bipolar electrode pairs, with vertical eye movements and blinks recorded from supraorbital and infraorbital electrodes at the left eye, and horizontal movements captured from electrodes at the outer canthi. All electrode impedances were maintained below 5 k$\Omega$ throughout the recording session. Data preprocessing was performed using the FieldTrip toolbox \cite{oostenveld2011fieldtrip}. The continuous EEG recordings were segmented into epochs time-locked to sentence onset, with each epoch spanning from 0 to 2,690 ms post-stimulus. Artifact correction was applied to epochs containing eye movements and muscle activity that exceeded a voltage threshold of 70 $\mu$V. All trials were retained for subsequent model training, validation, and testing.

This study focused on the alpha oscillations [8-13 Hz] in the EEG signals from 16 electrodes (Fz, FCz, FC1, FC2, Cz, C1, C2, CPz, CP1, CP2, Pz, P1, P2, POz, PO3, PO4), where inhibitory alpha activities are present in the data. The EEG signals were processed using a fifth-order Chebyshev Type II bandpass filter and subsequently downsampled to 200 Hz. Data from all EEG channels were normalized to ensure a mean of zero and a variance of one. To ensure that the data contained only stimulus-evoked neural activity, this study used EEG data time-locked to stimulus onset and epoched to 2.691 seconds, the minimum time required to complete a stimulus listening. Subsequently, discrete wavelet transform (DWT) \cite{shen2022eeg} was applied to the EEG signals to obtain a time-frequency representation that preserves 
both temporal and spectral information, which was then used as input to the model.

\subsubsection{Statistical Validation of the EEG Dataset}
To validate our EEG dataset and examine the effects of different listening conditions on alpha power, we conducted a repeated-measures ANOVA analysis followed by post-hoc comparisons. Alpha power (8-13 Hz) was calculated for each participant across the four listening conditions: clean, noisy, MMSE-enhanced, and Transformer-enhanced speech. For post-hoc pairwise comparisons, we used paired t-tests with false discovery rate (FDR) correction to control for multiple comparisons \cite{benjamini1995controlling}. This analysis was used to determine whether alpha power significantly differed between listening conditions, particularly between unprocessed noisy speech and clean/enhanced conditions. Statistical significance was set at $p < .05$ for all analyses. These statistical procedures provided a firm basis for validating the sensitivity of alpha oscillations as an objective measure of listening effort across different acoustic conditions in our large-scale dataset.

\subsection{Proposed Deep Learning Framework}

\begin{figure}[t!]
    \centering
    \includegraphics[width=\linewidth]{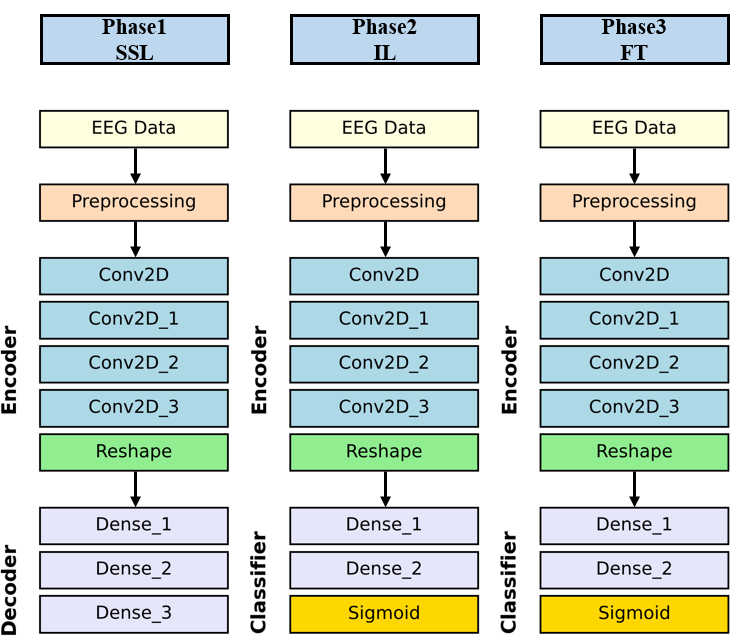}
    \caption
    {The architecture of EffortNet. 
    Phase 1(self-supervised learning): Unlabeled data from source domain; Phase 2(incremental learning): Labeled data from the source domain; Phase 3(fine-tuning): Labeled data from target domain.}
    \label{fig:fig1}
\end{figure}

This study proposed EffortNet, a three-phase training framework (see Fig. \ref{fig:fig1}) that integrates SSL, IL, and F-T to address the challenges of cross-subject EEG signal decoding. In Phase 1, a masking-based SSL approach was employed to learn generalizable feature representations from a large amount of unlabeled EEG data, reducing the reliance on labeled samples and enhancing cross-subject generalization. In Phase 2, the model was initialized with the encoder pretrained in Phase 1, and IL was incorporated to enable continual adaptation as subsequent subject data were sequentially introduced.
Finally, in Phase 3, the model was fine-tuned to optimize performance on the target domain data for personalized assessment.

\subsubsection{Phase 1: Self-Supervised Learning Pretraining on the Source Domain}

Given an unlabeled dataset $\mathcal{D}_{\text{src}}^{\text{u}}$ collected from 100 subjects in the source domain, we first pretrain an encoder $f_\theta$ and a decoder $g_\phi$ using a SSL objective~\cite{rafiei2022self,jia2025contrastive}, where $\theta$ and $\phi$ denote the trainable parameters of the encoder and decoder, respectively.

In SSL, a masked version $\tilde{X}$ of the input $X$ is provided to the model, and the model is trained to reconstruct the original signal. 
The mean squared error (MSE)~\cite{jia2025contrastive} is used as the loss function $\mathcal{L}(\cdot)$ in this phase to compare the reconstructed output with the original input.

The SSL loss in the source domain is defined as:
\begin{equation}
\mathcal{L}_{\text{SSL}}(\theta, \phi) = \mathcal{L}(g_\phi(f_\theta(\tilde{X})), X)
\end{equation}

We then optimize the encoder and decoder parameters by solving:
\begin{equation}
(\theta^*, \phi^*) = \arg\min_{\theta, \phi} \mathcal{L}_{\text{SSL}}(\theta, \phi)
\end{equation}

The resulting encoder parameters $\theta^*$ are then used to initialize the subsequent training phases.

\subsubsection{Phase 2: Incremental Learning with Replay in the Source Domain}

We adopt IL to enable continual adaptation to newly arriving data over time~\cite{zhou2024class,zhong2025replay}.
During this phase, the labeled dataset in the source domain is partitioned into $T = 100$  subsets, denoted as $\{ \mathcal{D}^{1}_{\text{src}}, \mathcal{D}^{2}_{\text{src}}, \dots, \mathcal{D}^{T}_{\text{src}} \}$, where each $\mathcal{D}^{t}_{\text{src}}$ corresponds to the EEG recordings from the $t$-th subject.

For each task $t$, the training data consists of two parts: the newly introduced labeled samples $(X^t, Y^t)$ specific to task $t$, and a subset of replayed labeled samples $(X', Y')$ drawn from previous tasks.

\begin{equation}
\mathcal{D}_{\text{src}}^t = \{(X^t, Y^t)\}, \quad \{(X', Y')\} \subseteq \bigcup_{i=1}^{t-1} \mathcal{D}_{\text{src}}^i
\end{equation}

The total loss at task $t$ for replay-based IL methods~\cite{zhong2025replay,deng2023centroid}, denoted as $\mathcal{L}_{\text{IL}}^t(\theta, \psi)$, is composed of two components: the loss on the current task data, $\mathcal{L}_{\text{current}}$, and the loss on replayed samples from previous tasks, $\mathcal{L}_{\text{replay}}$. 
The loss function used in this phase is the cross-entropy loss\cite{deng2023centroid}, denoted as $\mathcal{L}(\cdot)$.

A weighting hyperparameter $\lambda$ is used to balance the contribution of the replay term. This can be expressed as:

\begin{align}
\mathcal{L}_{\text{IL}}^t(\theta, \psi) = \ 
& \mathcal{L}_{\text{current}}(h_\psi(f_\theta(X^t)), Y^t) \nonumber \\
& + \lambda \cdot \mathcal{L}_{\text{replay}}(h_\psi(f_\theta(X')), Y')
\end{align}

In our implementation, we set the balancing hyperparameter to $\lambda = 1$.
In the first task, the encoder $f_\theta$ is initialized with parameters $\theta^*$ obtained from Phase~1. For each subsequent task $t > 1$, the encoder parameters $\theta^t$ are updated incrementally based on the parameters learned from the previous task, i.e., $\theta^{t-1}$. 
Each task $t$ uses a task-specific classifier $h_\psi$, which is further trained based on the parameters learned from the previous task $(t{-}1)$, resulting in the classifier parameters $\psi^t$ for the current task.

The optimization for each task can be formally expressed as:
\begin{equation}
(\theta^t, \psi^t) = \arg\min_{\theta, \psi} \mathcal{L}_{\text{IL}}^t(\theta, \psi), \quad \text{for } t = 1, \dots, T
\end{equation}

The resulting parameters \((\theta^T, \psi^T)\) from the source domain are subsequently used to initialize the encoder and classifier in the next phase, where we adapt the model to the target domain.

\subsubsection{Phase 3: Fine-Tuning on the Target Domain}

To enable knowledge transfer, we use the parameters $(\theta^T, \psi^T)$, corresponding to the encoder and classifier respectively, obtained from Phase~2 as initialization for fine-tuning in the target domain. Given a labeled dataset $\mathcal{D}_{\text{tgt}}^{\text{l}} = \{(X_{\text{tgt}}, Y_{\text{tgt}})\}$ in the target domain, we fine-tune\cite{li2021can} the model using supervised learning.

The supervised loss on the target domain, denoted as $\mathcal{L}(\cdot)$, is defined using the cross-entropy loss and is minimized during this phase:
\begin{equation}
\mathcal{L}_{\text{target}}(\theta, \psi) = \mathcal{L}(h_\psi(f_\theta(X_{\text{tgt}})), Y_{\text{tgt}})
\end{equation}

The model parameters are then optimized by solving:
\begin{equation}
(\theta^{\text{F-T}}, \psi^{\text{F-T}}) = \arg\min_{\theta, \psi} \mathcal{L}_{\text{target}}(\theta, \psi)
\end{equation}

This fine-tuning phase adapts the pretrained knowledge from the source domain to the distributional characteristics and label structure of the target domain. The resulting parameters $(\theta^{\text{F-T}}, \psi^{\text{F-T}})$ represent the final model used for inference on the target domain.

\subsubsection{Convolutional Neural Network Architecture}
The proposed deep learning framework was implemented using a convolutional neural network (CNN) architecture
\cite{zhi2023multi}. The CNN architecture consists of four convolutional layers designed to extract both spatial and temporal features from the input EEG data. The reshape layer transforms the multi-dimensional tensor output of the convolutional layers into one-dimensional data, which serves as the input to the dense layer. Subsequently, two fully connected dense layers map the extracted features into high-level representations, thereby capturing complex feature relationships. Finally, the Sigmoid activation function generates probability values for binary classification tasks.
\subsection{Training Setup}
\subsubsection{Training and Evaluation Setup}

The setup of this study follows a cross-subject training paradigm involving both a source and a target domain. The source domain consists of EEG data from 100 participants that were used to train the initial model. The target domain contains data from 20 unseen participants. For participant-specific adaptation, the model was fine-tuned using varying proportions of labeled EEG data from the target participants, with the remaining data used for evaluation.

Five-fold cross-validation was employed during the training process. Training was conducted using cross-entropy loss and the Adam optimizer (learning rate = $1 \times 10^{-4}$), with implementation details available on GitHub.

\subsubsection{Evaluation Metrics}
For alpha-band classification, accuracy was used as the primary metric, defined as:
\begin{align}
\text{Accuracy} = \frac{TP + TN}{TP + TN + FP + FN}
\end{align}
where TP, TN, FP, and FN denote true positives, true negatives, false positives, and false negatives, respectively \cite{hossin2015review}.

\section{RESULTS}
\subsection{Behavioral Results}
\begin{figure}[t!]
    \centering
    \includegraphics[width=\linewidth]{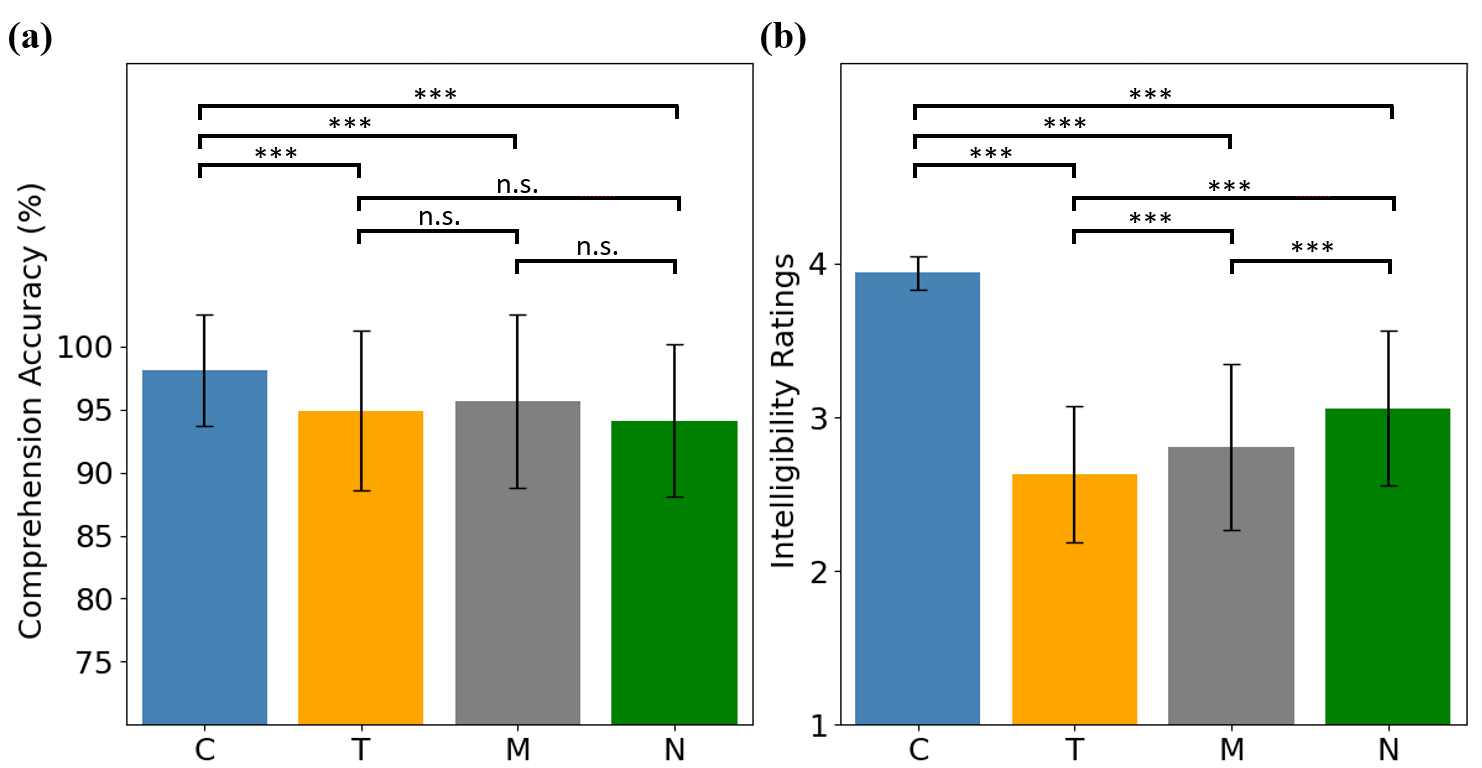}
\caption{Behavioral performance across the four listening conditions (C = Clean, N = Noisy, M = MMSE-enhanced, T = Transformer-enhanced). (a) Mean comprehension accuracy (in percent) for each condition. (b) Mean subjective intelligibility ratings (1–4 scale) for each condition. Error bars represent standard errors of the mean. Asterisks above brackets denote significant pairwise differences based on paired $t$-tests: $p < .05$ (*), $p < .01$ (**), $p < .001$ (***). n.s. denotes not significant.}
    \label{fig:fig20}
\end{figure}

Participants’ comprehension accuracy remained high across conditions (see Fig. \ref{fig:fig20} (a)), with mean accuracy of $98.11$\% ($SD = 4.45$) in the clean condition, $94.14$\% ($SD = 6.06$) in the noisy condition, $95.70$\% ($SD = 6.86$) in the MMSE-enhanced condition, and $94.90$\% ($SD = 6.33$) in the Transformer-enhanced condition. A one-way repeated-measures ANOVA revealed a significant main effect of condition on accuracy ($F_{3, 363} = 12.72, p < .001$). Paired $t$-tests showed that accuracy was significantly higher in the clean condition compared with the noisy ($t = 6.72, p < .001$), MMSE-enhanced ($t = 4.17, p < .001$), and Transformer-enhanced ($t = 5.24, p < .001$) conditions. No significant differences were found between the noisy and MMSE-enhanced ($t = -1.97, p = .307$), noisy and Transformer-enhanced ($t = -1.05, p = 1.000$), or between the two enhanced conditions ($t = 1.06, p = 1.000$).

Subjective intelligibility ratings are shown in Fig. \ref{fig:fig20} (b). The clean condition received the highest ratings ($M = 3.94, SD = 0.12$), followed by the noisy ($M = 3.06, SD = 0.50$), MMSE-enhanced ($M = 2.81, SD = 0.54$), and Transformer-enhanced ($M = 2.63, SD = 0.44$) conditions. A one-way repeated-measures ANOVA revealed a significant main effect of condition ($F_{3, 363} = 490.13, p < .001$). Paired $t$-tests indicated that all pairwise comparisons were significant: clean vs. noisy ($t = 20.50, p < .001$), clean vs. MMSE-enhanced ($t = 24.70, p < .001$), clean vs. Transformer-enhanced ($t = 35.00, p < .001$), noisy vs. MMSE-enhanced ($t = 9.35, p < .001$), noisy vs. Transformer-enhanced ($t = 12.90, p < .001$), and MMSE-enhanced vs. Transformer-enhanced ($t = 5.41, p < .001$).

\subsection{EEG Results: Repeated-measures ANOVA Analysis}
\begin{figure*}[t]
    \centering
    \includegraphics[width=\textwidth,keepaspectratio]{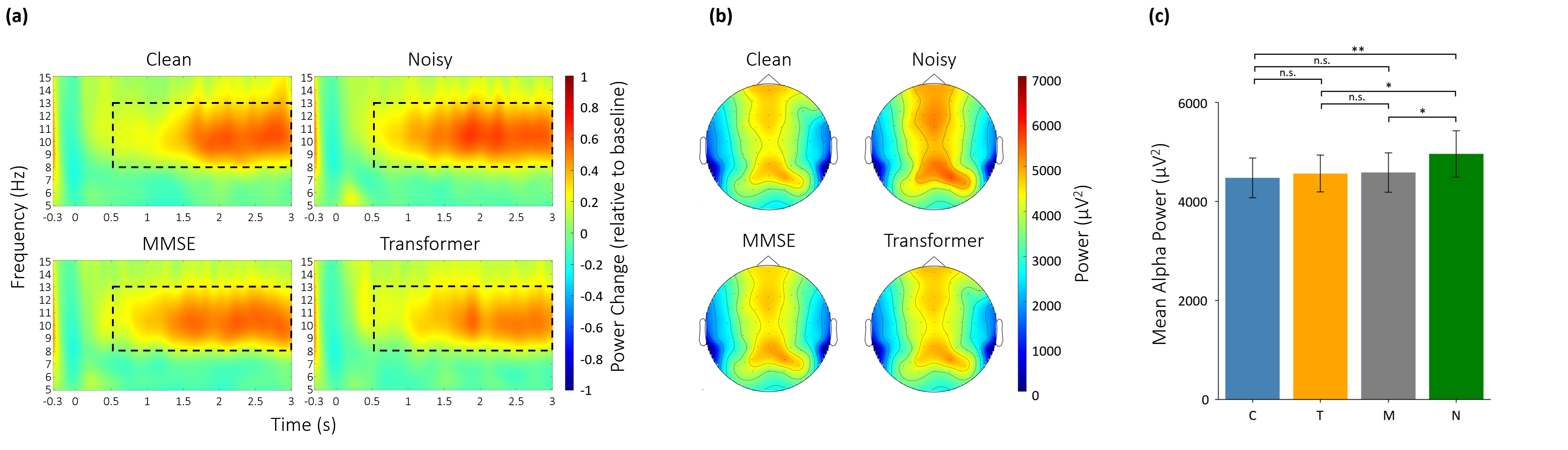}
    \caption{(a) Time-frequency representations across four listening conditions. Black dashed box highlights the alpha power (8–13 Hz) analysis window.
        (b) Topographic maps of absolute alpha power for each condition.
        (c) Mean alpha power across the four listening conditions (C = Clean, N = Noisy, M = MMSE-enhanced, T = Transformer-enhanced). Error bars: 95\% confidence interval. Asterisks above brackets: *$p < .05$, **$p < .01$, ***$p < .001$ (paired $t$ test); n.s., not significant.
    }
    \label{fig:fig8}
\end{figure*}

Fig. \ref{fig:fig8}(a) shows averaged time-frequency representations across 122 participants using 16 frontocentral and centroparietal channels (Fz, FC1, FCz, FC2, C1, Cz, C2, CP1, CPz, CP2, P1, Pz, P2, PO3, POz, and PO4). The analysis revealed increased alpha power (8–13 Hz) relative to baseline across all four listening conditions. The absolute alpha power data are presented as topographic maps in Fig. \ref{fig:fig8}(b), showing the spatial distribution averaged over 0–2.69 s for each condition.

Mean alpha power values (± SE) were 4478.8 ± 400.7 for clean, 4962.6 ± 468.2 for noisy, 4586.9 ± 398.8 for MMSE-enhanced, and 4564.6 ± 370.8 for Transformer-enhanced conditions (see Fig. \ref{fig:fig8}(a) and (b)).
A repeated-measures ANOVA revealed a significant main effect of listening condition on alpha power ($F_{(3, 363)}$
 = 5.99, $p < .01$, Greenhouse-Geisser corrected).
As shown in Fig. \ref{fig:fig8}(c), post-hoc analysis showed that alpha power during noisy condition was significantly higher compared with clean ($p < .01$, FDR-corrected), MMSE-enhanced ($p < .05$), and Transformer-enhanced ($p < .05$) conditions. No significant differences were found between clean and MMSE-enhanced, clean and Transformer-enhanced, or between MMSE-enhanced and Transformer-enhanced conditions ($ps > .05$).

\subsection{EEG Results: Deep Learning Analysis}
As illustrated in Fig. \ref{fig:fig8}(c), the noisy condition showed a significant increase in the alpha-band power compared with the clean condition. Therefore, our DL analysis first aimed to classify listening effort based on alpha power between these two conditions.

\subsubsection{Performance Comparisons with Baseline Models}

We adopted a commonly used ratio of training data at 0.8 as the baseline condition for training the classification model.
As shown in Fig. \ref{fig:fig18}, the EffortNet achieved a classification accuracy of $84.3$\% while the baseline CNN model and the STAnet achieved classification accuracies of $83.2$\% and $81.2$\%, respectively.  All three methods achieved average classification accuracies over $80$\%, confirming the effectiveness of DL-based methods for classification tasks on EEG signals.

 \subsubsection{Effect of Training Data Ratio on Model Performance}

\begin{figure}[t!]
    \centering
    \includegraphics[width=\linewidth]{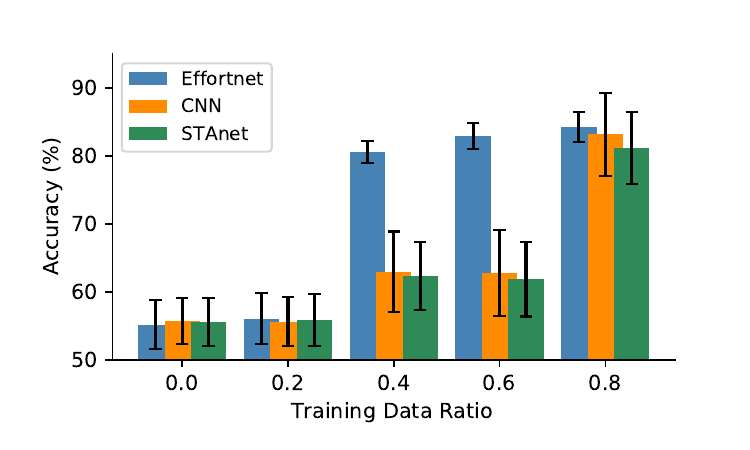}
\caption{ 
Classification performance comparisons between
accuracy and training data ratio.
}
  \label{fig:fig18}
\end{figure}

We further examined the effect of reduced training data on model performance, which was evaluated under various ratios of training data (see Fig. \ref{fig:fig18}). 

When the training ratio was 0.8, all models achieved high classification accuracy. EffortNet performed best at $84.3$\%, followed by CNN at $83.2$\%, and STAnet at $81.2$\%. At a training ratio of 0.6, EffortNet maintained strong performance at $82.9$\%, while both CNN and STAnet showed significant decreases to $62.7$\% and $61.9$\%, respectively. Similarly, with a training ratio of 0.4, EffortNet continued to demonstrate robust performance at $80.9$\%, substantially outperforming CNN ($62.3$\%) and STAnet ($61.1$\%).

As the training ratio decreased to 0.2, all models experienced considerable performance degradation. EffortNet's accuracy dropped to $56.1$\%, while CNN and STAnet showed comparable performance at $55.6$\% and $55.9$\%, respectively. Under the subject-independent condition (training ratio = 0.0), all models performed at near-chance levels: EffortNet at $55.2$\%, CNN at $55.7$\%, and STAnet at $55.6$\%.

\subsubsection{Performance on Unseen Participants}

\begin{figure*}[t]
    \centering
    \includegraphics[width=\textwidth,keepaspectratio]{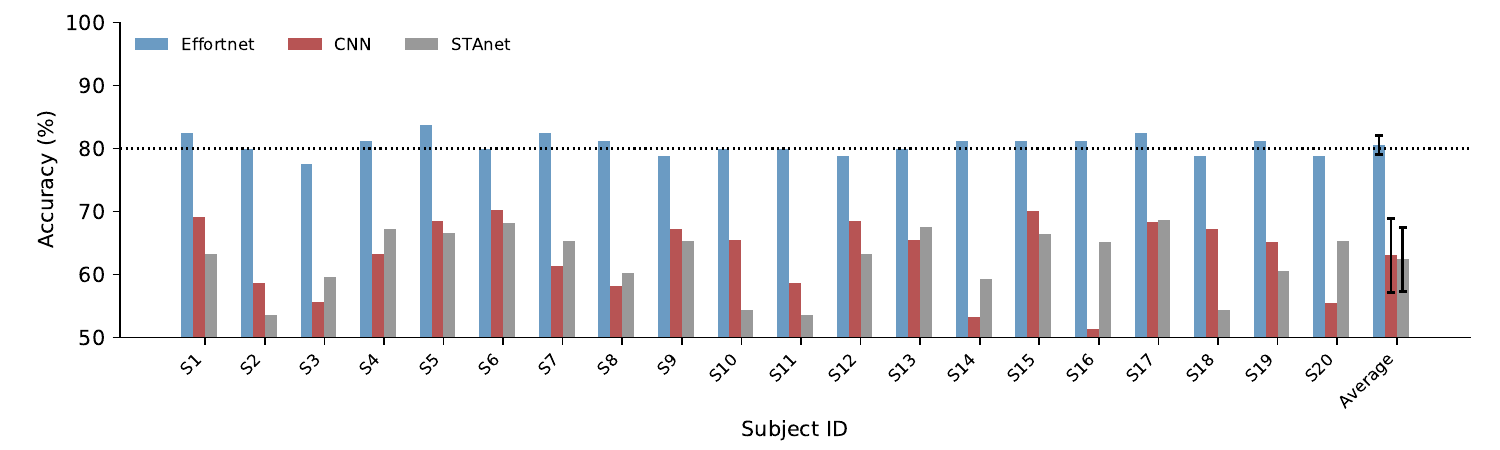}
\caption{
Comparisons of the model performance on 20 unseen participants with a training data ratio at 0.4.
The horizontal dotted line represents a reference point of
accuracy at $80$\%.}
  \label{fig:fig11}
\end{figure*}

We subsequently examined the generalizability of EffortNet by evaluating its performance on the data from 20 previously unseen participants through fine-tuning with the training data ratio at 0.4. As shown in Fig. \ref{fig:fig11}, the EffortNet achieved an average accuracy of $80.9$\% ($SD = 3.0$). However, the baseline CNN model and the STAnet showed average accuracies of $62.4$\% ($SD = 6.5$) and $61.2$\% ($SD = 5.3$), respectively.

\subsubsection{Ablation Analysis on Model Performance}
\begin{figure*}[t]
    \centering
    \includegraphics[width=\textwidth,keepaspectratio]{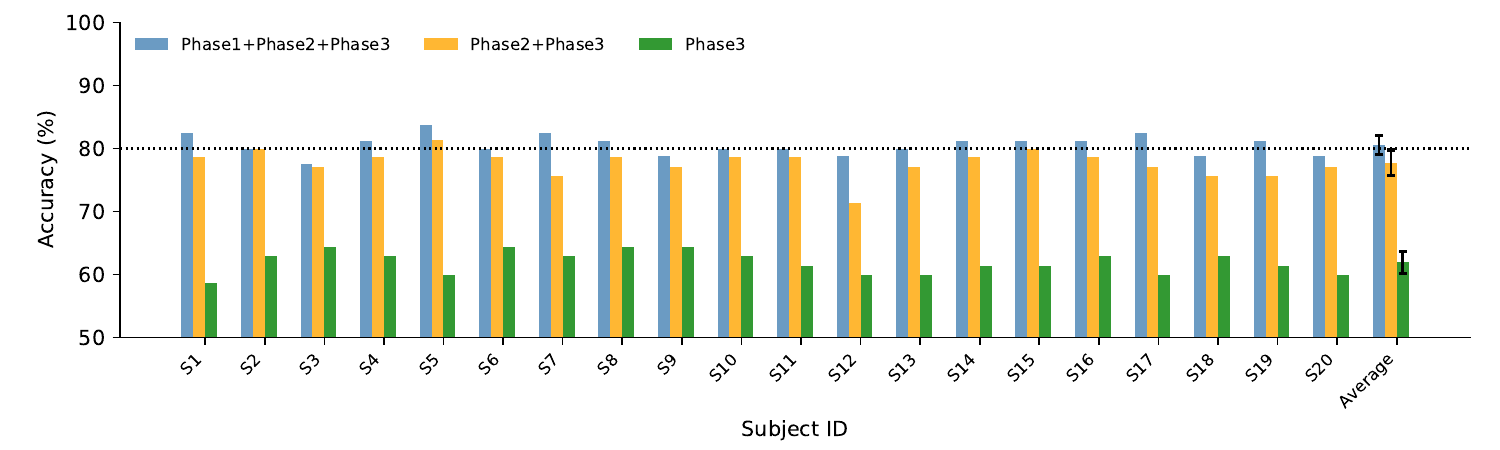}
\caption{
Classification accuracies on the 20 unseen participants under different phase configurations of the EffortNet model in the ablation analysis. The horizontal dotted line represents a reference point of accuracy at $80$\%.}
  \label{fig:fig12}
\end{figure*}

As shown in Fig. \ref{fig:fig12}, the complete three-phase model achieved $80.9$\% ($SD = 3.0$) accuracy. Next, the model with a Phase 2 and 3 configuration achieved an accuracy of $77.7$\% ($SD = 2.1$). Finally, the model with only a Phase 3 configuration showed an accuracy of $61.9$\% ($SD = 1.7$). Overall, the complete three-phase model demonstrated the best performance among all competing configurations, with Phase 2 (IL) contributing to the model's performance most.

\subsection{Probability-based Metric using Alpha-band Power to Evaluate Speech Enhancement Effectiveness}

As the EffortNet has demonstrated robust classification accuracies on alpha power of the clean and noisy conditions, we subsequently utilized this model to evaluate speech enhancement technologies by having the model assess EEG responses from the MMSE- and Transformer-enhanced conditions as more probable to be responses from the clean or noisy conditions.

\begin{figure}[t]
    \centering
    \includegraphics[width=\linewidth]{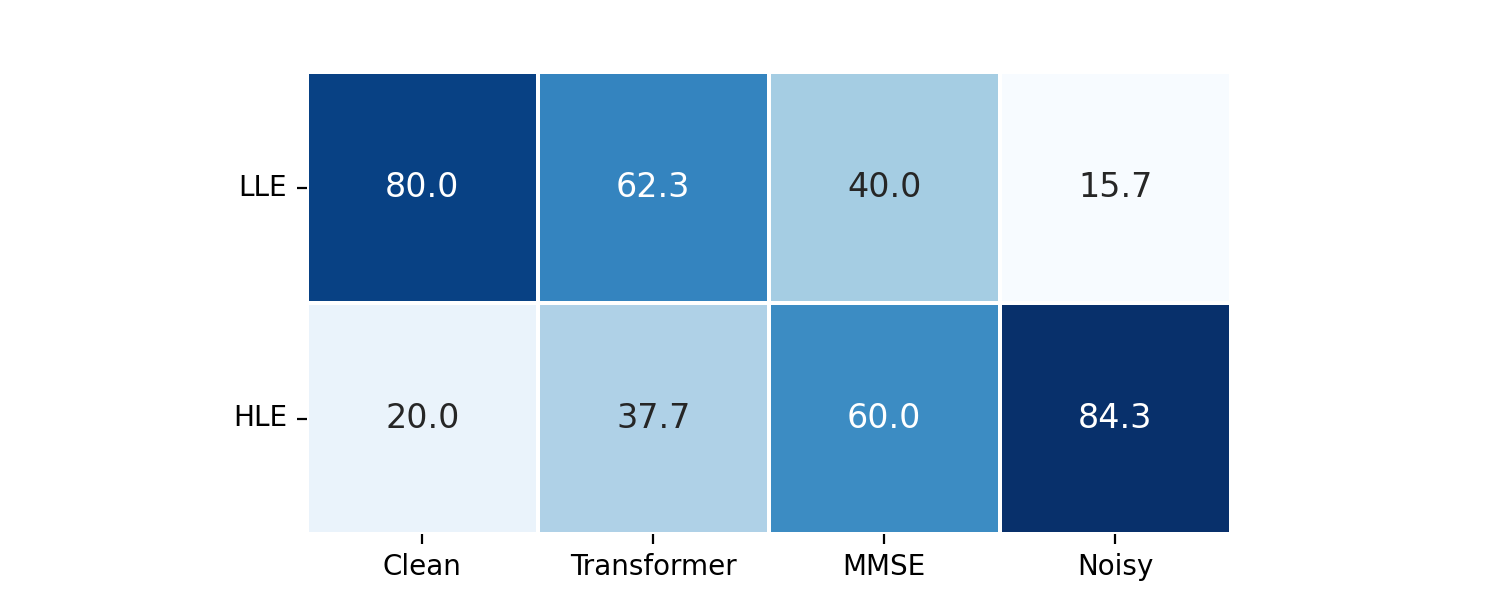}
\caption{EffortNet confusion matrix showing predicted probabilities of classifying alpha activity from four speech conditions as low listening effort(LLE) or high listening effort(HLE).}
  \label{fig:fig16}
\end{figure}

In this scenario, we defined the model's predicted results as low listening effort (LLE) when an alpha power response was classified as being elicited by clean speech whereas the results were defined as high listening effort (HLE) when an alpha power response was classified as being elicited by noisy speech. As shown in Fig. \ref{fig:fig16}, EffortNet was used as a speech enhancement evaluator. The probabilities of alpha power responses being predicted as LLE were $80$\%, $62.3$\%, $40.0$\%, and $15.7$\% for clean, Transformer-enhanced, MMSE-enhanced, and noisy conditions, respectively. This decreasing trend suggests that the Transformer-based enhancement algorithm produced speech signals that elicited brain responses more similar to clean speech than the MMSE-based algorithm, with both enhancement methods producing responses more similar to clean speech than the noisy condition.

\begin{figure}[t]
    \centering
    \includegraphics[width=\linewidth]{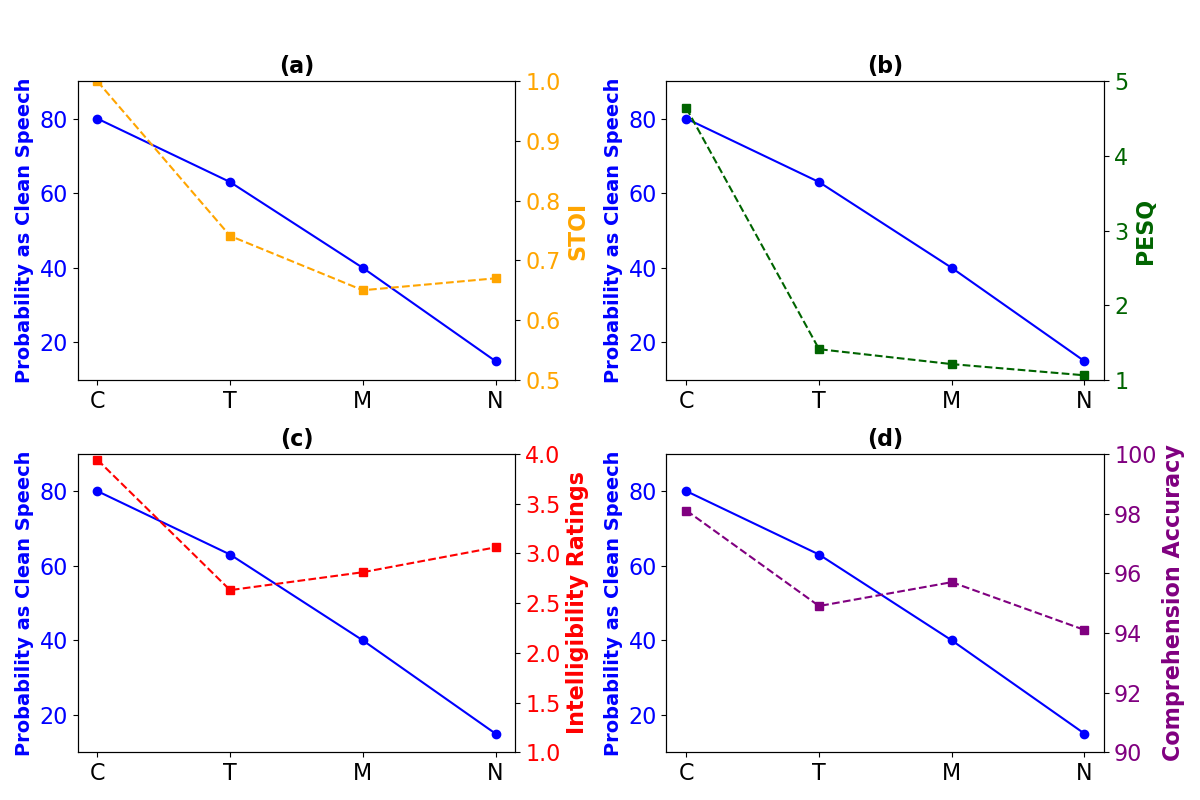}
\caption{Comparison of the EffortNet-based alpha-band metric for listening effort with standard evaluation measures in four listening conditions (C = Clean, T = Transformer-enhanced, M = MMSE-enhanced, N = Noisy): (a) STOI scores, (b) PESQ scores, (c) human intelligibility ratings, and (d) comprehension accuracy.}

  \label{fig:fig19}
\end{figure}

We further compared these alpha-based classification results with those from established automatic metrics and behavioral measures, as presented in Fig. \ref{fig:fig19}. The objective metrics showed trends consistent with our neural measures, where the alpha-band metric (probability of alpha activity resembling clean speech patterns) indicated that higher probabilities correspond to lower listening effort. From Fig. \ref{fig:fig19}(a), the STOI scores (measuring speech intelligibility) under the acoustic conditions clean, Transformer, MMSE, and noisy were 1.00, 0.74, 0.65, and 0.64, respectively. Similarly, in Fig. \ref{fig:fig19}(b), the PESQ scores (measuring speech quality) were 4.64, 1.41, 1.21, and 1.06, respectively. These results indicate a decreasing trend that closely aligns with our EEG-based evaluation, suggesting that neural responses may reflect objective speech measures.

In contrast, behavioral measures exhibited divergent trends.
In Fig. \ref{fig:fig19}(c), the subjective intelligibility ratings under clean, Transformer, MMSE, and noisy conditions were 3.94, 2.63, 2.81, and 3.06, respectively, where enhanced speech received lower ratings compared with noisy speech, and the Transformer-enhanced speech was rated lower than the MMSE-enhanced speech. In Fig. \ref{fig:fig19}(d), participants achieved comprehension accuracies of $98.11$\%, $94.90$\%, $95.70$\%, and $94.14$\% under clean, Transformer, MMSE, and noisy conditions, with participants performing worst in the noisy condition. 
This discrepancy underscores the complementary role of neural metrics in capturing cognitive responses that may not be reflected by subjective assessments or behavioral performance alone.

\section{DISCUSSIONS}
\subsection{Alpha Oscillations and Listening Effort}

The comparison between our behavioral and EEG findings reveals a notable divergence that aligns with the measurement inconsistencies highlighted in the previous research. While comprehension accuracy showed minimal differences between noisy and enhanced conditions, alpha power, which indexes listening effort, demonstrated significant reductions in both enhancement conditions compared with the noisy condition. This dissociation illustrates the fundamental challenge in assessing listening effort.

Our behavioral results indicate that participants maintained high comprehension accuracy across all conditions, with only the clean condition showing significantly higher performance. This ceiling effect in comprehension is not uncommon in speech perception studies using sentence material \cite{hsin2023speech,hsin2025exploring}, as participants can often compensate for acoustic degradation through increased cognitive effort. However, this compensation comes at a cost that behavioral measures may not fully capture. Similarly, subjective intelligibility ratings showed a different pattern, with participants rating noisy speech significantly higher than both enhanced conditions, despite comparable comprehension performance.

In contrast to these behavioral results, our EEG findings based on a large-scale dataset from healthy young participants revealed that alpha power was significantly higher during the noisy condition compared to both clean and enhanced conditions. This increased alpha power during noisy speech processing reflects greater listening effort, supporting previous findings that alpha oscillations serve as a mechanism for auditory selective inhibition during challenging listening conditions \cite{strauss2014cortical,weisz2011alpha} and that increased power reflects the suppression of task-irrelevant information \cite{wostmann2016spatiotemporal,wostmann2019alpha}.

The observed dissociation between behavioral measures and neurophysiological indicators substantiates recent research showing limited correlations between different measures of listening effort \cite{alhanbali2019measures,francis2020listening,shields2023exploring}. Different measurement domains likely tap into distinct dimensions of listening effort rather than reflecting a unified construct. Our results are consistent with previous findings that demonstrated that physiological measures can detect changes in listening effort even when behavioral performance remains stable\cite{mcgarrigle2017measuring,mcgarrigle2017pupillometry, hsin2023speech}.

Importantly, our EEG results provide neurophysiological evidence that both enhancement algorithms (MMSE and Transformer) effectively reduced listening effort compared with noisy speech, despite participants' subjective impressions. This finding aligns with previous work by Slugocki et al. \cite{slugocki2024alpha,slugocki2024using}, who showed that effective hearing aid algorithms can significantly reduce alpha power during speech-in-noise tasks. The absence of significant differences in alpha power between clean speech and enhanced conditions further suggests that both enhancement algorithms successfully restored neural processing to levels comparable to clean speech processing.

These findings underscore the value of alpha oscillations as an objective biomarker for quantifying listening effort during speech comprehension. While behavioral measures may be influenced by conscious perception, task strategy, or motivation, alpha oscillations provide a more direct window into the neural processes underlying listening effort. This neurophysiological approach offers a promising solution to the measurement challenges, which we have further addressed by utilizing DL-based techniques at individual levels.

\subsection{Decoding Individual Listening Effort}
While our statistical analyses confirmed significant differences in alpha power across listening conditions at the group level, our DL-based approach offers critical advantages for practical, individual-level assessment of listening effort. The proposed EffortNet model enables accurate classification of listening effort for clean or noisy speech with accuracies exceeding $80$\%, even with limited training data, addressing the challenge of substantial inter-individual variability in EEG signals.

Our DL framework integrates three complementary learning paradigms that collectively overcome key challenges in EEG classification. SSL effectively leverages unlabeled data through augmentation techniques, reducing dependence on extensive labeled data while improving model robustness against noise and signal variability. IL addresses the non-independent and identically distributed nature of multi-subject EEG data by progressively adapting to subject-specific distributions, significantly outperforming traditional batch training methods. Our ablation results confirm this advantage, showing that the IL method contributed most significantly to model performance. 
Finally, F-T enables efficient adaptation to new individuals by leveraging knowledge from existing data, thereby reducing the need for extensive new data collection and enhancing practicality in real-world applications. The superior performance of EffortNet compared with conventional CNN and STAnet models demonstrates its potential for deployment with limited calibration data. This efficiency is particularly valuable in clinical and commercial applications where user convenience and rapid adaptation are essential.

\subsection{Personalized Assessment of Speech Enhancement Effectiveness}

This study demonstrates how alpha oscillations, when analyzed through DL approaches, can provide an objective, personalized evaluation of speech enhancement technologies based on listening effort. By defining EEG responses elicited by clean speech as indicative of LLE and those evoked by noisy speech as HLE, we established an evaluation framework to quantify the effectiveness of speech enhancement technologies at the individual level.

The EffortNet classifier revealed a clear pattern across listening conditions: $80$\% of EEG responses to clean speech were classified as LLE, compared to $62.3$\% for Transformer-enhanced speech, $40$\% for MMSE-enhanced speech, and only $15.7$\% for noisy speech. This gradient demonstrates the classifier's sensitivity to varying degrees of listening effort across different acoustic conditions. Importantly, the Transformer-based enhancement algorithm produced speech signals that elicited brain responses more similar to clean speech than the MMSE-based algorithm, with both enhancement methods producing responses more similar to clean speech than the noisy condition.

Notably, our neural measure-based assessments aligned well with established objective metrics but diverged from subjective ratings. The STOI scores (measuring speech intelligibility) under clean, Transformer-enhanced, MMSE-enhanced, and noisy conditions were 1.00, 0.74, 0.65, and 0.64, respectively, while PESQ scores (measuring speech quality) were 4.64, 1.41, 1.21, and 1.06. These objective metrics showed a decreasing trend consistent with our alpha-based classification. In contrast, subjective intelligibility ratings exhibited an opposite pattern, with participants rating enhanced speech lower than noisy speech (3.94, 2.63, 2.81, and 3.06 for clean, Transformer-enhanced, MMSE-enhanced, and noisy conditions, respectively).

This discrepancy between neural measures and subjective ratings echoes the measurement inconsistencies we highlighted earlier \cite{mcgarrigle2014listening, alhanbali2019measures, shields2023exploring}. While participants subjectively rated enhanced speech lower than noisy speech, their neural responses indicated reduced listening effort with enhanced speech. This finding aligns with previous research showing that subjective perceptions may not reflect underlying cognitive processes \cite{francis2020listening, pichora2016hearing}, and highlights the value of neurophysiological measures in capturing aspects of listening effort that may not be consciously accessible. Further analysis revealed a parallel trend between alpha power and objective speech metrics (STOI and PESQ), with lower alpha power (indicating reduced listening effort) associated with higher speech quality and intelligibility scores. This trend provides some validation for our DL classification approach and demonstrates the potential of alpha oscillations as an objective biomarker for evaluating speech enhancement technologies.

The practical implications of this approach are significant. Traditional evaluation of hearing assistance technologies often relies on subjective ratings or speech recognition performance, which may not capture the cognitive effort involved in listening. Our framework offers a neurophysiologically-based method to evaluate the actual cognitive impact of speech enhancement technologies, enabling more personalized and effective interventions for individuals with hearing difficulties. This approach could significantly influence the development and selection of hearing assistance technologies, focusing on reducing cognitive load during listening comprehension.

\section{Conclusion}

This study addresses a critical challenge in hearing healthcare—developing reliable, objective measurements for evaluating listening effort and assistive hearing technologies. Our work responds to the call for improved methodologies to assess listening effort\cite{pichora2016hearing}, a multidimensional construct impacting quality of life, particularly in aging populations at risk for dementia. We have demonstrated that alpha oscillations serve as a robust biomarker of listening effort, with increased power during noisy speech processing reflecting greater cognitive demand. While statistical analyses revealed significant group-level differences, this study's innovation lies in the personalized assessment enabled by our deep learning framework. The EffortNet model addresses three fundamental challenges: it provides objective assessment aligning with established metrics while capturing cognitive processes not reflected in behavioral measures; it handles individual variability through integrating SSL, IL, and F-T, maintaining high classification accuracy with limited training data; and it enables practical application requiring minimal user calibration.

Our probability-based metric revealed that Transformer-enhanced speech elicits neural responses more similar to clean speech than MMSE-enhanced speech. While subjective ratings favored unprocessed noisy speech, alpha power measures demonstrated reduced effort with enhanced speech, highlighting the disconnect between subjective perception and neural processing. The parallel trend between alpha power and objective speech metrics validates our approach and demonstrates EEG's potential as a complementary evaluation tool. By focusing on cognitive impact, our framework offers comprehensive assessment of how these technologies reduce listening effort. As global populations age and hearing loss becomes prevalent, developing effective interventions to mitigate cognitive decline is imperative. This study provides methodological advances and practical tools for personalized hearing healthcare, potentially influencing technology development focused on reducing listening effort.

In summary, this study contributes through: (1) establishing a large-scale EEG dataset for speech comprehension under various acoustic conditions; (2) demonstrating alpha oscillations' utility as a personalized assessment tool; and (3) validating a DL technique addressing individual variability while minimizing data requirements. These advances represent important steps toward bridging laboratory insights and clinical applications.

\section*{Acknowledgment}
The authors sincerely thank Dr. Syu-Siang Wang for his insightful comments and suggestions on the early draft of the manuscript. We also thank Wen-Yuan Ting for helpful suggestions on the manuscript. Finally, we extend our heartfelt appreciation to all study participants for their essential contributions.

\section*{Data Availability}
The codes and EEG dataset are available on GitHub.\\ (https://github.com/JohnSung0501/EEG-Effort-Classification)

\section*{References}

\bibliographystyle{IEEEtran}
{
\bibliography{REF_TNSRE}
}


\vfill

\end{document}